\documentclass[12pt]{article}
\begin{document}
\title{The Dark Energy Universe}
\author{B.G. Sidharth\\
International Institute for Applicable Mathematics \& Information Sciences\\
Hyderabad (India) \& Udine (Italy)\\
B.M. Birla Science Centre, Adarsh Nagar, Hyderabad - 500 063
(India)}
\date{}
\maketitle
\begin{abstract}
Some seventy five years ago, the concept of dark matter was
introduced by Zwicky to explain the anomaly of galactic rotation
curves, though there is no clue to its identity or existence to
date. In 1997, the author had introduced a model of the universe
which went diametrically opposite to the existing paradigm  which
was a dark matter assisted decelarating universe. The new model
introduces a dark energy driven accelarating universe though with a
small cosmological constant. The very next year this new picture was
confirmed by the Supernova observations of Perlmutter, Riess and
Schmidt. These astronomers got the 2011 Nobel Prize for this
dramatic observation. All this is discussed briefly, including the
fact that dark energy may obviate the need for dark matter.
\end{abstract}
\section{Introduction}
By the end of the last century, the Big Bang Model had been worked
out. It contained a huge amount of unobserved, hypothesized "matter"
of a new kind - dark matter. This was postulated as long back as the
1930s to explain the fact that the velocity curves of the stars in
the galaxies did not fall off, as they should. Instead they
flattened out, suggesting that the galaxies contained some
undetected and therefore non-luminous or {\bf dark matter}. The
identity of this dark matter has been a matter of guess work,
though. It could consist of Weakly Interacting Massive Particles
(WIMPS) or Super Symmetric partners of existing particles. Or heavy
neutrinos or monopoles or unobserved brown dwarf stars and so on.\\
In fact Prof. Abdus Salam speculated some two decades ago
\cite{salamnap} "And now we come upon the question of dark matter
which is one of the open problems of cosmology. This is a problem
which was speculated upon by Zwicky fifty years ago. He showed that
visible matter of the mass of the galaxies in the Coma cluster was
inadequate to keep the galactic cluster bound. Oort claimed that the
mass necessary to keep our own galaxy together was at least three
times that concentrated into observable stars. And this in turn has
emerged as a central problem
of cosmology.\\
"You see there is the matter which we see in our galaxy. This is
what we suspect from the spiral character of the galaxy keeping it
together. And there is dark matter which is not seen at all by any
means whatsoever. Now the question is what does the dark matter
consist of? This is what we suspect should be there to keep the
galaxy bound. And so three times the mass of the matter here in our
galaxy should be around in the form of the invisible matter. This is
one of the
speculations."\\
The universe in this picture, contained enough of the mysterious
dark matter to halt the expansion and eventually trigger the next
collapse. It must be mentioned that the latest WMAP survey
\cite{science2}, in a model dependent result indicates that as much
as twenty three percent of the Universe is made up of dark matter,
though there is no definite observational confirmation
of its existence.\\
That is, the Universe would expand
up to a point and then collapse.\\
There still were several subtler problems to be addressed. One was
the famous {\bf horizon problem}. To put it simply, the Big Bang was
an uncontrolled or random event and so, different parts of the
Universe in different directions were disconnected at the very
earliest stage and even today, light would not have had enough time
to connect them. So they need not be the same. Observation however
shows that the Universe is by and large uniform, rather like people
in different countries showing the same habits or dress. That would
not be possible without some form of faster than light
intercommunication which would violate Einstein's Special
Theory of Relativity.\\
The next problem was that according to Einstein, due to the material
content in the Universe, space should be curved whereas the Universe
appears to be {\bf flat}.\\
There were other problems as well. For example astronomers predicted
that there should be {\bf monopoles} that is, simply put, either
only North magnetic poles or only South magnetic poles, unlike the
North South combined magnetic poles we encounter. Such monopoles
have
failed to show up even after seventy five years.\\
Some of these problems as we noted, were sought to be explained by
what has been called {\bf inflationary cosmology} whereby, early on,
just after
the Big Bang the explosion was super fast \cite{zee,lindepl82}.\\
What would happen in this case is, that different parts of the
Universe, which could not be accessible by light, would now get
connected. At the same time, the super fast expansion in the initial
stages would smoothen out any distortion or curvature effects in
space,
leading to a flat Universe and in the process also eliminate the monopoles.\\
Nevertheless, inflation theory has its problems. It does not seem to
explain the cosmological constant observed since. Further, this
theory seems to imply that the fluctuations it produces should
continue to indefinite distances. Observation seems to imply the
contrary.\\
One other feature that has been studied in detail over the past few
decades is that of {\bf structure formation} in the Universe. To put
it simply, why is the Universe not a uniform spread of matter and
radiation? On the contrary it is very lumpy with planets, stars,
galaxies and so on, with a lot of space separating these objects.
This has been explained in terms of fluctuations in density, that
is, accidentally more matter being present in a given region.
Gravitation would then draw in even more matter and so on. These
fluctuations would also cause the cosmic background radiation to be
non uniform or anisotropic. Such anisotropies are in fact being
observed. But this is not the end of the story. The galaxies seem to
be arranged along two dimensional structures and filaments with huge
separating
voids.\\
From 1997, the conventional wisdom of cosmology that had concretized
from the mid sixties onwards, began to be challenged. It had been
believed that the density of the Universe is near its critical
value, separating eternal expansion and ultimate contraction, while
the nuances of the dark matter theories were being fine tuned. But
that year, the author proposed a contra view, which we will examine.
\section{Cosmology}
To proceed, as there are $N \sim 10^{80}$ such particles in the
Universe, we get, consistently,
\begin{equation}
Nm = M\label{3e1}
\end{equation}
where $M$ is the mass of the Universe. It must be remembered that
the energy of gravitational interaction between the particles is
very much insignificant compared to the above electromagnetic considerations.\\
In the following we will use $N$ as the sole cosmological parameter.\\
We next invoke the well known relation
\cite{bgsfluc,nottalefractal,hayakawa}
\begin{equation}
R \approx \frac{GM}{c^2}\label{3e2}
\end{equation}
where $M$ can be obtained from (\ref{3e1}). We can arrive at
(\ref{3e2}) in different ways. For example, in a uniformly expanding
Friedman Universe, we have
$$\dot{R}^{2} = 8 \pi G\rho R^2/3$$
In the above if we substitute $\dot{R} = c$ at $R$, the radius of
the universe, we
get (\ref{3e2}). Another proof will be given later in Section 3.10.\\
We now use the fact that given $N$ particles, the
(Gaussian)fluctuation in the particle number is of the order
$\sqrt{N}$\cite{hayakawa,huang,ijmpa,ijtp,bgsfqp,bgsmg8}, while a
typical time interval for the fluctuations is $\sim \hbar/mc^2$, the
Compton time, the fuzzy interval within which there is no meaningful
physics as argued by Dirac and in greater detail by Wigner and
Salecker. So particles are created and destroyed - but the ultimate
result is that $\sqrt{N}$ particles are created just as this is the
nett displacement in a random walk of unit step. So we have,
\begin{equation}
\frac{dN}{dt} = \frac{\sqrt{N}}{\tau}\label{3ex}
\end{equation}
whence on integration we get, (remembering that we are almost in the
continuum region that is, $\tau \sim 10^{-23}sec \approx 0$),
\begin{equation}
T = \frac{\hbar}{mc^2} \sqrt{N}\label{3e3}
\end{equation}
We can easily verify that the equation (\ref{3e3}) is indeed
satisfied where $T$ is the age of the Universe. Next by
differentiating (\ref{3e2}) with respect to $t$ we get
\begin{equation}
\frac{dR}{dt} \approx HR\label{3e4}
\end{equation}
where $H$ in (\ref{3e4}) can be identified with the Hubble Constant,
and using (\ref{3e2}) is given by,
\begin{equation}
H = \frac{Gm^3c}{\hbar^2}\label{3e5}
\end{equation}
Equation (\ref{3e1}), (\ref{3e2}) and (\ref{3e3}) show that in this
formulation, the correct mass, radius, Hubble constant and age of
the Universe can be deduced given $N$, the number of particles, as
the sole cosmological or large scale parameter. We observe that at
this stage we are not invoking any particular dynamics - the
expansion is due to the random creation of particles from the ZPF
background. Equation (\ref{3e5}) can be written as
\begin{equation}
m \approx \left(\frac{H\hbar^2}{Gc}\right)^{\frac{1}{3}}\label{3e6}
\end{equation}
Equation (\ref{3e6}) has been empirically known as an "accidental"
or "mysterious" relation. As observed by Weinberg \cite{weinberggc},
this is unexplained: it relates a single cosmological parameter $H$
to constants from microphysics. We will touch upon this micro-macro
nexus again. In our formulation, equation (\ref{3e6}) is no longer a
mysterious coincidence but
rather a consequence of the theory.\\
As (\ref{3e5}) and (\ref{3e4}) are not exact equations but rather,
order of magnitude relations, it follows, on differentiating
(\ref{3e4}) that a small cosmological constant $\wedge$ is allowed
such that
$$\wedge \leq 0 (H^2)$$
This is consistent with observation and shows that $\wedge$ is very
small $--$ this has been a puzzle, the so called cosmological
constant problem alluded to, because in conventional theory, it
turns out to be huge \cite{weinbergprl}. But it poses no problem in
this formulation. This is because of the characterization of the ZPF
as independent and primary in our formulation this being the
mysterious dark energy. Otherwise we would encounter the
cosmological constant problem of Weinberg: a $\wedge$ that is some
$10^{120}$ orders of magnitude of observable values!\\
To proceed we observe that because of the fluctuation of $\sim
\sqrt{N}$ (due to the ZPF), there is an excess electrical potential
energy of the electron, which in fact we identify as its inertial
energy. That is \cite{ijmpa,hayakawa},
$$\sqrt{N} e^2/R \approx mc^2.$$
On using (\ref{3e2}) in the above, we recover the well known
Gravitation-Electromagnetism ratio viz.,
\begin{equation}
e^2/Gm^2 \sim \sqrt{N} \approx 10^{40}\label{3e7}
\end{equation}
or without using (\ref{3e2}), we get, instead, the well known so
called Weyl-Eddington formula,
\begin{equation}
R = \sqrt{N}l\label{3e8}
\end{equation}
(It appears that (\ref{3e8}) was first noticed by H. Weyl
\cite{singh}). Infact (\ref{3e8}) is the spatial counterpart of
(\ref{3e3}). If we combine (\ref{3e8}) and (\ref{3e2}), we get,
\begin{equation}
\frac{Gm}{lc^2} = \frac{1}{\sqrt{N}} \propto T^{-1}\label{3e9}
\end{equation}
where in (\ref{3e9}), we have used (\ref{3e3}). Following Dirac
(cf.also \cite{melnikov}) we treat $G$ as the variable, rather than
the quantities $m, l, c \,\mbox{and}\, \hbar$ which we will call
micro physical constants because of their central role
in atomic (and sub atomic) physics.\\
Next if we use $G$ from (\ref{3e9}) in (\ref{3e5}), we can see that
\begin{equation}
H = \frac{c}{l} \quad \frac{1}{\sqrt{N}}\label{3e10}
\end{equation}
Thus apart from the fact that $H$ has the same inverse time
dependance on $T$ as $G$, (\ref{3e10}) shows that given the
microphysical constants, and
$N$, we can deduce the Hubble Constant also, as from (\ref{3e10}) or (\ref{3e5}).\\
Using (\ref{3e1}) and (\ref{3e2}), we can now deduce that
\begin{equation}
\rho \approx \frac{m}{l^3} \quad \frac{1}{\sqrt{N}}\label{3e11}
\end{equation}
Next (\ref{3e8}) and (\ref{3e3}) give,
\begin{equation}
R = cT\label{3e12}
\end{equation}
Equations (\ref{3e11}) and (\ref{3e12}) are consistent with observation.\\
Finally, we observe that using $M,G \mbox{and} H$ from the above, we
get
$$M = \frac{c^3}{GH}$$
This relation is required in the Friedman model of the expanding
Universe (and the Steady State model too). In fact if we use in this
relation, the expression,
$$H = c/R$$
which follows from (\ref{3e10}) and (\ref{3e8}), then we recover
(\ref{3e2}). We will be repeatedly using these relations in the sequel.\\
 As we saw the above model predicts a dark
energy driven ever expanding and accelerating Universe with a small
cosmological constant while the density keeps decreasing. Moreover
mysterious large number relations like (\ref{3e5}), (\ref{3e11}) or
(\ref{3e8}) which were considered to be miraculous accidents now
follow from the underlying theory. This seemed to go against the
accepted idea that the density of the Universe equalled the critical
density required for closure and that aided by dark matter, the
Universe was decelerating.\\
However, as noted, from 1998 onwards, following the work of {\bf
Perlmutter}, {\bf Schmidt} and {\bf Riess}, these otherwise apparently heretic conclusions have been vindicated.\\
It may be mentioned that the observational evidence for an
accelerating Universe was the American Association for Advancement
of Science's Breakthrough of the Year, 1998 while the evidence for
nearly seventy five percent of the Universe being Dark Energy, based
on the Wilkinson Microwave Anisotropy Probe (WMAP) and the Sloan Sky
Digital Survey was the Breakthrough of the Year, 2003
\cite{science1,science2}. The trio got the 2011 Nobel for Physics.\\
The zero point field or dark energy has another signature. It causes for example the Lamb Shift, which for a Hydrogen atom is $1 MHz$ and is ubiquitous. This gives us an approximate idea of the strength of the signal. So we should expect a uniform Cosmic Radio Background of roughly $1 MHz$ to about 1 metre, remembering the various dissipative processes that exist in space.\\
NASA's ARCADE, balloon borne experiment detected a mysterious isotropic radio radiation (or hiss) six times as powerful as expected, but not from any specific radio sources. This is consistent with a power law and is precisely in the wavelength region of $10cm$ to 1 metre. This otherwise inexplicable radio wave background could well be the above signature.
\section{Discussion}
1. We observe that in the above scheme if the Compton time $\tau \to
\tau_P$, we recover the Prigogine Cosmology \cite{prig,tryon}. In
this case there is a {\bf phase transition} in the background ZPF or
Quantum Vacuum or Dark Energy and Planck scale particles are
produced.\\
On the other hand if $\tau \to 0$ (that is we return to point
spacetime), we recover the Standard Big Bang picture. But it must be
emphasized that in neither of these two special cases can we recover
the various so called Large Number coincidences for example
Equations like (\ref{3e3}) or (\ref{3e5}) or (\ref{3e7}) or
(\ref{3e8}).\\
2. The above ideas lead to an important characterization of
gravitation. This also explains why it has not been possible to
unify gravitation with other interactions, despite nearly a century
of effort.\\
Gravitation is the only interaction that could not be satisfactorily
unified with the other fundamental interactions. The starting point
has been a diffusion equation
$$| \Delta x|^2 = < \Delta x^2 > = \nu \cdot \Delta t$$
\begin{equation}
\nu = \hbar/m, \nu \approx l v\label{2e3}
\end{equation}
This way we could explain a process similar to the formation of
Benard cells \cite{tduniv,prig} -- there would be sudden formation
of the ``cells" from the background dark energy, each at the Planck
Scale, which is the smallest physical scale. These in turn would be
the underpinning for
spacetime.\\
We could consider an array of $N$ such Planckian cells
\cite{bgsijtp}. This would be described by
\begin{equation}
r  = \sqrt{N \Delta x^2}\label{4De1d}
\end{equation}
\begin{equation}
ka^2 \equiv k \Delta x^2 = \frac{1}{2}  k_B T\label{4De2d}
\end{equation}
where $k_B$ is the Boltzmann constant, $T$ the temperature, $r$ the
extent  and $k$ is the spring constant given by
\begin{equation}
\omega_0^2 = \frac{k}{m}\label{4De3d}
\end{equation}
\begin{equation}
\omega = \left(\frac{k}{m}a^2\right)^{\frac{1}{2}} \frac{1}{r} =
\omega_0 \frac{a}{r}\label{4De4d}
\end{equation}
We now identify the particles or cells with \index{Planck}Planck
\index{mass}masses and set $\Delta x \equiv a = l_P$, the
\index{Planck}Planck length. It may be immediately observed that use
of (\ref{4De3d}) and (\ref{4De2d}) gives $k_B T \sim m_P c^2$, which
ofcourse agrees with the temperature of a \index{black hole}black
hole of \index{Planck}Planck \index{mass}mass. Indeed, Rosen
\cite{rosen} had shown that a \index{Planck}Planck \index{mass}mass
particle at the \index{Planck scale}Planck scale can be considered
to be a \index{Universe}Universe in itself with a Schwarzchild
radius equalling the Planck length. We also use the fact alluded to
that a typical elementary particle like the \index{pion}pion can be
considered to be the result of $n \sim 10^{40}$ \index{Planck}Planck
\index{mass}masses.\\
Using this in (\ref{4De1d}), we get $r \sim l$, the \index{pion}pion
\index{Compton wavelength}Compton wavelength as required. Whence the
pion mass is given by
$$m = m_P/\sqrt{n}$$
which of course is correct, with the choice of $n$. This can be
described by
\begin{equation}
l = \sqrt{n} l_P, \, \tau = \sqrt{n} \tau_P,\label{3e31}
\end{equation}
$$l^2_P = \frac{\hbar}{m_P} \tau_P$$
The last equation is the analogue of the diffusion process seen,
which is in fact the underpinning for particles, except that this
time we have the same Brownian process operating from the Planck
scale to the Compton scale (Cf. also \cite{bgsfpl152002,cu}).\\
We now use the well known result alluded to that the individual
minimal oscillators are black holes or mini Universes as shown by
Rosen \cite{rosen}. So using the Beckenstein temperature formula for
these primordial black holes \cite{ruffinizang}, that is
$$kT = \frac{\hbar c^3}{8\pi Gm}$$
we can show that
\begin{equation}
Gm^2 \sim \hbar c\label{4e4}
\end{equation}
We can easily verify that (\ref{4e4}) leads to the value $m \sim
10^{-5}gms$. In deducing (\ref{4e4}) we have used the typical
expressions for the frequency as the inverse of the time - the
Compton time in this case and similarly the expression for the
Compton length. However it must be reiterated that no specific
values
for $l$ or $m$ were considered in the deduction of (\ref{4e4}).\\
We now make two interesting comments. Cercignani and co-workers have
shown \cite{cer1,cer2} that when the gravitational energy becomes of
the order of the electromagnetic energy in the case of the Zero
Point oscillators, that is
\begin{equation}
\frac{G\hbar^2 \omega^3}{c^5} \sim \hbar \omega\label{4e5}
\end{equation}
then this defines a threshold frequency $\omega_{max}$ above which
the oscillations become chaotic. In other words, for meaningful
physics we require that
$$\omega \leq \omega_{max}.$$
Secondly as we can see from the parallel but unrelated theory of
phonons \cite{huang,reif}, which are also bosonic oscillators, we
deduce a maximal frequency given by
\begin{equation}
\omega^2_{max} = \frac{c^2}{l^2}\label{4e6}
\end{equation}
In (\ref{4e6}) $c$ is, in the particular case of phonons, the
velocity of propagation, that is the velocity of sound, whereas in
our case this velocity is that of light. Frequencies greater than
$\omega_{max}$ in (\ref{4e6}) are again meaningless.
We can easily verify that using (\ref{4e5}) in (\ref{4e6}) gives back (\ref{4e4}).\\
In other words, gravitation shows up as the residual energy from the
formation of the particles in the universe via Planck scales (Benard
like) cells.\\
3. It has been mentioned that despite nearly 75 years of search,
Dark Matter has not been found. More recently there is evidence
against the existence of Dark Matter or its previous models. The
latest LHC results for example seem to rule out SUSY.\\
On the other hand our formulation obviates the need for Dark Matter.
This follows from an equation like (\ref{3e9}) which shows a
gravitational constant decreasing with time. Starting from here it
is possible to deduce not just the anomalous rotation curves of
galaxies which was the starting point for Dark Matter; but also we
could deduce all the known standard results of General Relativity
like the precession of the perihelion of mercury, the bending of
light, the progressive shortening of the time period of binary
pulsars and so on (Cf.ref.\cite{tduniv}).\\
4. {\bf Epilogue}: The idea of a perfect vacuum began to get frayed
in the 19th century. In the 20th century with the advent of Quantum
Theory the concept of a Quantum Vacuum came into being. This Quantum
Vacuum is seething with energy and activity, and it is there
everywhere. With this background we can see the following:\\
Around 1997 I had put forward a radically different model. In this,
there wasn't any \index{Big Bang}Big Bang, with matter and energy
being created instantaneously. Rather the universe is permeated by
an energy field of a kind familiar to modern physicists. The point
is, that according to Quantum theory which is undoubtedly one of the
great intellectual triumphs of the twentieth century, all our
measurements, and that includes measurements of energy, are at best
approximate. There is always a residual error. This leads to what
physicists call a ubiquitous \index{Zero Point Field}Zero Point
Field or \index{Quantum Vacuum}Quantum Vacuum. We will return to
this ``\index{Dark Energy}Dark Energy" soon. Out of such a ghost
background or all pervading energy field, particles are created in a
totally random manner, a process that keeps continuing. However,
much of the matter was created in a fraction of a second. There is
no ``Big Bang" singularity, though, which had posed Wheeler's
greatest problem of physics. The contents of this paper went
diametrically opposite to accepted ideas, that the universe,
dominated by dark matter was actually decelarating. Rather, driven
by dark energy, the universe would be expanding and accelerating,
though slowly. I was quite sure that this paper would be rejected
outright by any reputable scientific journal. So I presented these
ideas at the prestigious \index{Marcell Grossmann meet}Marcell
Grossmann meet in Jerusalem and another International Conference on
Quantum Physics. But, not giving into pessimism, I shot off the
paper to a standard International journal, anyway. To my great
surprise,
it was accepted immediately!\\
There is a further cosmic foot print of this model: a residual
miniscule energy in the \index{Cosmic Microwave Background}Cosmic
Microwave Background, less than a billion billion billion billionth
of the energy of an electron. Latest data has confirmed the presence
of such an energy. All this is in the spirit of the manifest
universe springing out of an unmanifest background, as described in
the \index{Bhagvad Gita}Bhagvad Gita. There are
several interesting consequences.\\
Firstly it is possible to theoretically estimate the size and age of
the universe and also deduce a number of very interesting
interrelationships between several physical quantities like the
charge of the electron, the mass of elementary particles, the
gravitational constant, the number of particles in the universe and
so on. One such, connecting the gravitational constant and the mass
of an elementary particle with the expansion of the universe was
dubbed as inexplicable by Nobel Laureate \index{Steven
Weinberg}Steven Weinberg. But on the whole these intriguing
interrelationships have been considered by
most scientists to be miraculous coincidences.\\
With one exception. The well known Nobel Prize winning physicist
\index{Paul Dirac}Paul Dirac sought to find an underlying reason to
explain what would otherwise pass off as a series of inexplicable
accidents. In this model, there is a departure from previous
theories including the fact that some supposedly constant quantities
like the universal constant of gravitation are actually varying very
slowly with time. Interestingly latest observations seem to point
the finger in this direction.\\
However my model is somewhat different and deduces these mysterious
relations. Further, it sticks its neck out in predicting that the
universe is not only expanding, but also accelerating as it does so.
This went against all known wisdom. Shortly thereafter from 1998
astronomers like \index{Perlmutter}Perlmutter and
\index{Kirshner}Kirshner began to publish observations which
confirmed exactly such a behavior. These shocking results have since
been reconfirmed. The universe had taken
a U Turn.\\
When questioned several astronomers in 1998 confided to me that the
observations were wrong! After the expansion was reconfirmed, some
became cautious. Let us wait and see. At the same time, some rushed
back to their desks and tried to rework their calculations. The
other matter was, what force could cause the accelerated expansion?
The answer would be, some new and inexplicable form of energy, as
suggested by me. \index{Dark Energy}Dark Energy. Later the presence
of dark energy was confirmed by the Wilkinson Microwave Probe
(\index{WMAP}WMAP) and the \index{Sloane Digital Sky Survey}Sloane
Digital Sky Survey. Both these findings were declared by the
prestigious journal Science as breakthroughs of the respective years.\\
The accelerated expansion of the universe and the possibility that
supposedly eternally constant quantities could vary, has been the
new paradigm gifted to science, a parting gift by the departing
millennium.\\
A 2000 article in the \index{Scientific American}Scientific American
observed, ``In recent years the field of cosmology has gone through
a radical upheaval. New discoveries have challenged long held
theories about the evolution of the Universe... Now that observers
have made a strong case for cosmic acceleration, theorists must
explain it.... If the recent turmoil is anything to go by, we had
better keep our options
open."\\
On the other hand, an article in \index{Physics World}Physics World
in the same year noted , ``A revolution is taking place in
cosmology. New ideas are usurping traditional notions about the
composition of the Universe, the relationship between geometry and
destiny, and Einstein's greatest blunder."\\
It is this greatest blunder of Einstein which got the Nobel Prize
for Physics in 2011 for three US astronomers, Perlmutter, Reiss and
Schmidt who observed the accelerated expansion of the universe in
1988.

\end{document}